\documentclass[twocolumn,showpacs,preprintnumbers,amsmath,amssymb]{revtex4}
\usepackage{graphicx}

\newcommand{\ie}{{\em i.e.~}}

\newcommand{\ea}{{~\it et al.}~}

\begin{document}

\title{Phenomenological model for a novel melt-freeze phase of sliding bilayers}

\author{Trieu Mai}
\affiliation{Department of Physics, University of California, Santa Cruz, CA 95064.}

\date{\today}

\begin{abstract}
Simulations show that sliding bilayers of colloidal particles can
exhibit a new phase, the ``melt-freeze'' phase, where the layers
stochastically alternate between solidlike and liquidlike states.
We introduce a mean field phenomenological model with two order
parameters to understand the interplay of two adjacent layers while
the system is in this remarkable phase. Predictions from our numerical
simulations of a system in the melt-freeze phase include the tendency
of two adjacent layers to be in opposite states (solid and liquid)
and the difference between the fluctuation of the order parameter
in one layer while the other layer is in the same phase compared
to the fluctuation while the other layer is in the opposite phase.  We
expect this behavior to be seen in future simulations and experiments.
\end{abstract}


\maketitle 

\section{Introduction} 

General theories of nonequilibrium phenomena are still elusive and
current research in nonequilibrium physics is directed on a
phenomenon-by-phenomenon basis. One important problem is the
shear-induced melting (and/or freezing) of a solid. The shear flow
of a solid is widely studied experimentally~\cite{persson} due to
its importance in material science. Being a standard example of the
rich field of nonequilibrium phase transitions, shear-induced melting
is also widely studied theoretically on systems as diverse as vortex
lattices to suspended colloidal particles
~\cite{persson,ackerson,stevens,Das1}.

For clarity, we will focus on a specific system of colloidal particles
on two dimensional sheets~\cite{ackerson}. An isolated sheet at low
enough temperature will have colloidal particles in an ordered
phase. When two such sheets are driven across each other, one sheet
subjects stress on the other (and vice versa) and at high enough
stresses, could induce melting~\cite{Das1}. At large enough driving
forces, the sheets are occasionally found to re-order to a re-entrant
solid phase displaying, the phenomenon of shear-induced
freezing~\cite{lahiri}.

Recently, numerical simulations of sliding bilayers by
Das\ea~\cite{Das1,Das2} discovered a new ``melt-freeze'' phase beyond
the standard shear-induced phases. Das\ea performed Brownian dynamics
simulations of particles on two adjacent monolayers in the manner
described in the previous paragraph. Earlier simulations usually
contain one monolayer driven across a fixed substrate~\cite{Das1}
which is applicable when one layer is much stiffer than the other.
However the simulations of Das\ea addressed the situation in which
two comparably soft layers are driven across each other and thus
the behavior of the particles on {\it both} layers becomes important.
The relevant parameters in their simulations included the strength
of the driving force, the magnitude of the coupling between particles
on differing sheets, and the noise amplitude. Changing these
parameters displays rich and unexpected behavior.

Das\ea found that at fixed interlayer coupling and noise strength,
the sytem undergoes different phases as the driving force is varied.
For small driving forces, the two monolayers remain in the ordered
phase and simply creep past one other. The same qualitative behavior
is seen for very large driving forces. However, at intermediate
driving forces, the two monolayers undergo a melt-freeze phase
in which both layers stochastically disorder (melt) and order
(freeze). As the interlayer coupling increases, the melt-freeze
phase persists over a larger range of driving forces. For a fixed
driving force (in which the system is in the melt-freeze phase),
increasing the coupling also increases the amount of time a layer
spends in the disordered state.

In addition to the simulation of particles on layers, Das\ea
introduced a simple mean field phenomenological model~\cite{Das1}.
This phenomenological model contains one order parameter and a
strain variable. Like the spatially dependent system, the mean field
model has three simulation parameters (noise strength, coupling
strength, and driving force). Remarkably, this simple model contains
much of the rich behavior of the more detailed particle simulations.
However, the interplay between the two layers cannot be studied by
this model with only one order parameter. The simple model also
clearly connects the melt-freeze phase with stochastic resonance
phenomena~\cite{Das1,mcnamara}. 

In this paper, we closely follow but extend the phenomenology
in~\cite{Das1} to include two order parameters $\rho_1$ and $\rho_2$
in the same spirit that Das\ea's particle simulations~\cite{Das1,Das2}
generalized the conventional models of one sheet sliding over a
fixed substrate. This natural extension qualitatively recovers
much of the behavior seen in the one order parameter phenomenological
model (therefore also in the particle model). In addition, we observe
two new features involving how one order parameter affects the free
energy landscape of the other. The first feature is the tendency
of the two order parameters to be in different phases when the
coupling is finite, \ie when $\rho_1$ is in the ordered phase,
$\rho_2$ is likely to be in the disordered phase. The second feature
involves the fluctuation of the order parameters. When $\rho_1$ and
$\rho_2$ are out of phase, their fluctuations are smaller than when
they are in the same phase. These features are compared to the
simulation of particles in~\cite{Das1,Das2}.

In the next section, we describe our two parameter Landau model and
the Langevin-Ginzburg dynamics. Section III contains the results
from our simulations. As in~\cite{Das1,Das2}, the results include
driving force dependence and coupling strength dependence.

\section{Model Free Energy and Dynamics}

The order parameter in our mean field model represents the structure
factor for the colloidal particles in a given sheet~\cite{order}.
The model free energy must represent a system with two phases
(crystal and liquid) and contain a first order transition between
the phases. A simple effective Landau free energy with these
considerations is the simple double well potential used in
Ref.~\cite{Das1,Das2},
\begin{equation}
V(\rho) = \frac{\alpha}{2}\rho^2-\frac{\beta}{3}\rho^3+\frac{\gamma}{4}\rho^4,
\label{2well}
\end{equation}
where $\alpha, \beta,$ and $\gamma$ are all positive.
The values are chosen so that there are two phases (two minima) and
the stable phase is the crystalline one when there is no stress.

Modelling the dynamically driven phase transitions requires a strain
variable $\theta$, as in~\cite{Das1,Das2}. Since stress is induced
by a periodic crystal, the strain variable must be periodic (we use
the range $[0,1)$). A liquid does not induce stress on a driven
solid sheet. Using these properties of strain, we model the part
of the free energy with the strain variable to be~\cite{strain}
\begin{equation}
W(\rho_1,\rho_2,\theta) = \frac{c}{2}\rho_1^2\rho_2^2[1-cos(2\pi\theta)],
\label{freestrain}
\end{equation}
where $c$ is a coupling strength. The total model free energy includes
both order parameters in a free energy landscape of Eq.(\ref{2well})
and the strain variable and coupling of Eq.(\ref{freestrain}),
\begin{equation}
F(\rho_1,\rho_2,\theta) = V(\rho_1)+V(\rho_2)+W(\rho_1,\rho_2,\theta).
\label{free}
\end{equation}

\begin{figure}
\begin{center}
\includegraphics[width=3.25in]{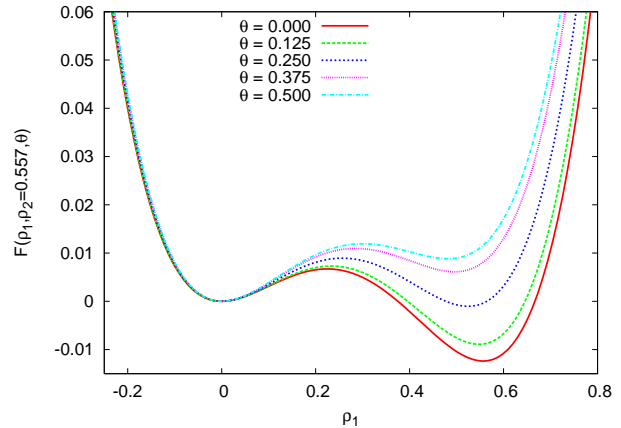}
\caption{
The free energy for one order parameter $\rho_1$ while the other
order parameter $\rho_2$ is in the ordered state. The different
plots represent different values of $\theta$ from 0.0 (no stress)
to 0.5 (maximal stress). The coupling $c$ is set to 0.25 here and
$\alpha=1,\beta=6.25$, and $\gamma=8$.}
\label{landscape}
\end{center}
\end{figure}
Figure \ref{landscape} shows the free energy for an order parameter
while the other is in the ordered state. Since the other sheet is
ordered, it can induce stress. As stress increases, shown by a
larger $\theta$, the ordered phase becomes less stable and the
system is driven towards the liquid phase. The changing landscape
shown in figure~\ref{landscape} shows how the simple free energy
of Eq.(\ref{free}) can represent shear-induced melting.

Eq.(\ref{free}) displays the equilibrium conditions in which the
order parameters exist. For these driven systems, the dynamics
are also important. In this paper, we use overdamped Langevin
dynamics of the form
\begin{eqnarray}
\dot{\rho}_1 & = & -\frac{1}{\Gamma_\rho}\frac{\partial F}{\partial\rho_1} 
+ \eta_\rho, \nonumber \\
\dot{\rho}_2 & = & -\frac{1}{\Gamma_\rho}\frac{\partial F}{\partial\rho_2} 
+ \eta_\rho, \nonumber \\
\dot{\theta} & = & -\frac{1}{\Gamma_\theta}\frac{\partial F}{\partial\theta} 
+ d + \eta_\theta.
\label{dyneq}
\end{eqnarray}
The damping coefficients $\Gamma$'s are set to unity, $d$
represents a constant driving force for the strain, and the $\eta$'s
represent noise. We use Gaussian noise with zero mean $<\eta> = 0$
and variance $D$, $<\eta(t)\eta(t^\prime)> = 2D\delta(t-t^\prime)$.
The amount of noise $D$ is related to the temperature through the
fluctuation-dissipation theorem.  The strain variable and the order
parameters have noise of equal amplitudes in our simulations.

Eq.(\ref{free}) and Eqs.(\ref{dyneq}) describe the free energy and
dynamics of our phenomenological model. In the next section, we
briefly describe the numerical implementation of these equations
and the results of our simulations.

\section{Results}

We numerically integrate the Langevin-Ginzburg equations,
Eqs.(\ref{dyneq}), using the fourth order Runge-Kutta algorithm.
For all simulations, the stepsize is $0.001$ and the integrator is
run for $\sim 10^6$ transient steps which are discarded. We also
set $\alpha=1, \beta=6.25$, and $\gamma=8$ for all simulations. As
in~\cite{Das1,Das2}, the adjustable parameters in the model are the
strength of the constant driving force $d$, the coupling strength
$c$, and the noise amplitude $D$. Below, we show how the system
behaves by varying these parameters independently. To study the
variation of the driving force and the coupling, we keep the noise
amplitude fixed at $D = 0.004$. We observe that this noise strength
is insufficient to drive the system to the disordered phase from
the ordered phase for very long times when stress is absent.

\subsection{Driving Force Dependence}

\begin{figure}
\begin{center}
\includegraphics[width=3in]{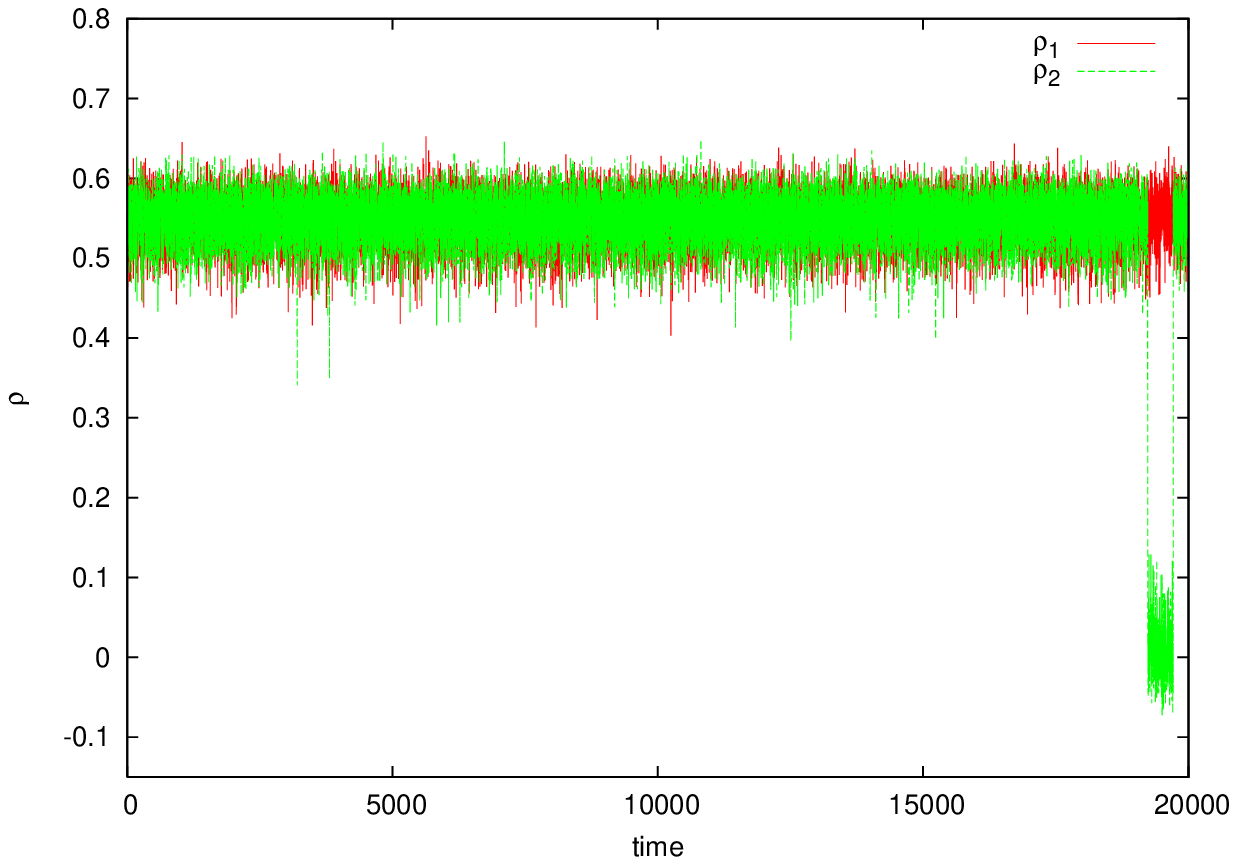} \\
\includegraphics[width=3in]{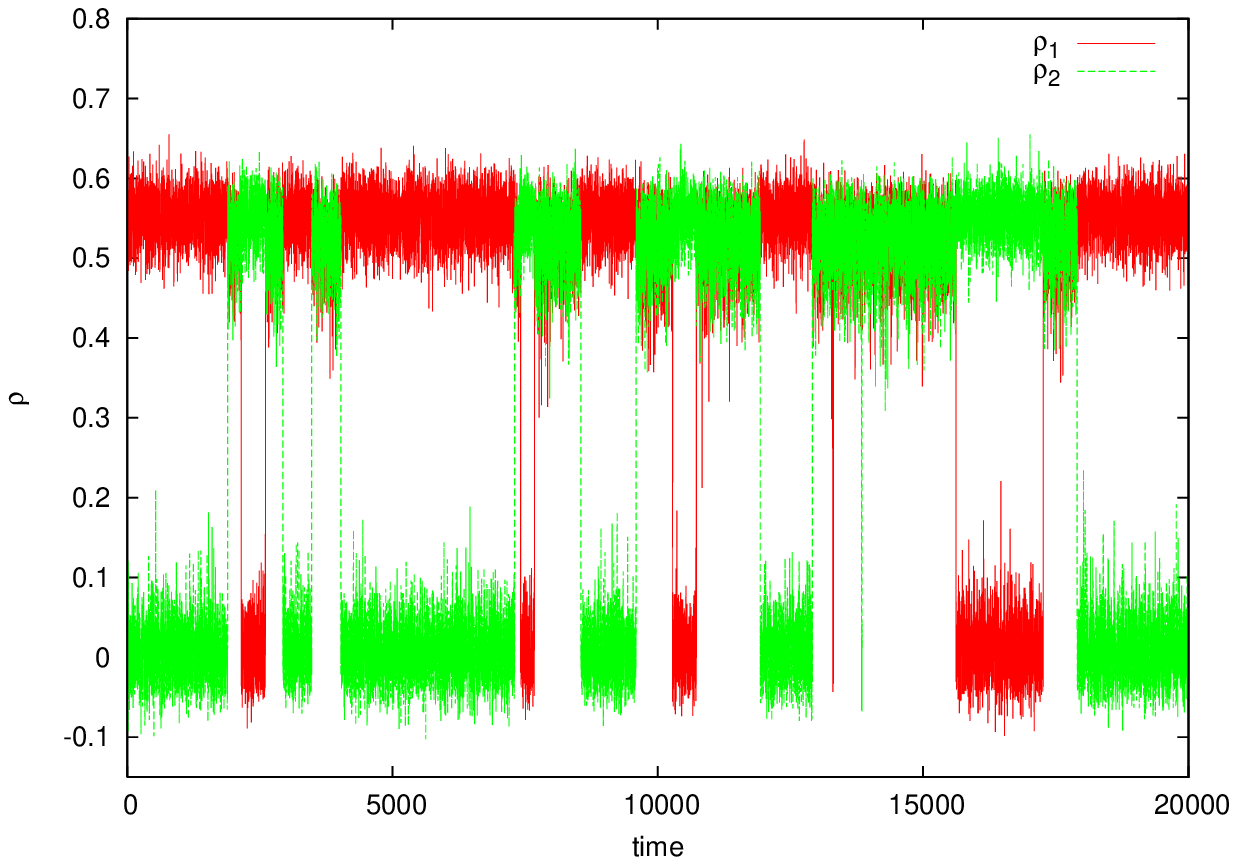} \\
\includegraphics[width=3in]{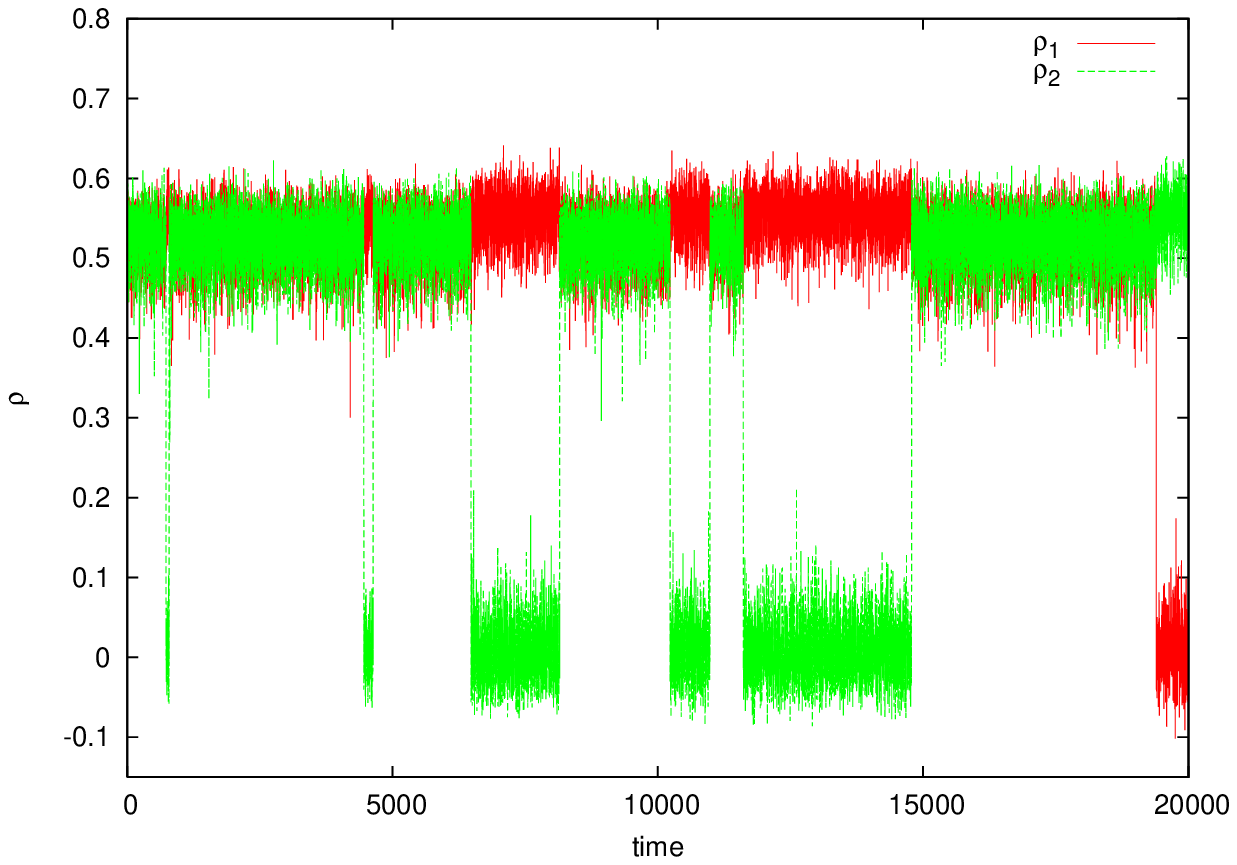} 
\caption{The order parameter vs. time for three different driving
forces $d = 0.025, 0.1,$ and $10.0$ respectively from top to bottom.
The coupling $c=0.25$ and noise strength $D=0.004$ are kept fixed.}
\label{drive}
\end{center}
\end{figure}

\begin{figure}
\begin{center}
\includegraphics[width=3.25in]{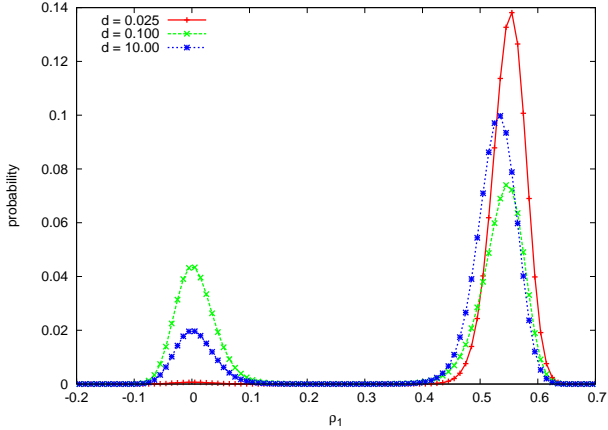} 
\caption{The distribution of $\rho_1$ taken over $10^{11}$
steps for $d = 0.025, 0.1,$ and $10.0$. The coupling $c=0.25$ and
noise strength $D=0.004$ are kept fixed.}
\label{prob_drive}
\end{center}
\end{figure}

As described in the introduction and in~\cite{Das1,Das2}, the system
stays essentially crystalline when the driving force is below a
threshold value or above another threshold. Only at intermediate
driving forces is the melt-freeze phase seen. In figure~\ref{drive}
we show $\rho_1$ and $\rho_2$ versus time at constant coupling, $c
= 0.25$. The unit of time in figure~\ref{drive} and figure~\ref{coup}
represents $2\times 10^4$ timesteps. Figure~\ref{prob_drive} shows
the distribution of $\rho_1$ (the distribution of $\rho_2$ is the
same after averaging over long enough time due to the symmetry
between the two in the free energy).

\begin{figure}
\begin{center}
\includegraphics[width=3.25in]{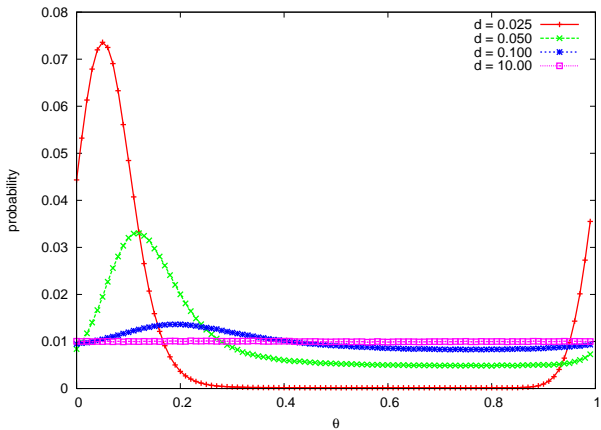}
\caption{The distribution of $\theta$ taken over $10^{11}$ steps
for $d = 0.025, 0.05, 0.1,$ and $10.0$. The coupling $c=0.25$ and
noise strength $D=0.004$ are kept fixed. $\theta$ is periodic in
the range [0,1).}
\label{strain_drive}
\end{center}
\end{figure}

To understand this behavior qualitatively, we discuss the role of
the strain variable in determining the phase of the system. The
driving force tilts the cosine potential and drives the strain
variable in a preferred direction. At very small driving forces,
the strain variable is rarely able to leave the minimum in which
it started. As seen from figure~\ref{landscape}, when $\theta$
is near this minimum the crystal well remains stable and $\rho$
does not cross the barrier. For very large driving forces the cosine
potential is not seen and the distribution of $\theta$ is essentially
flat and does not spend enough time near its peak to allow $\rho$
to cross over to the liquid well. In between these two limits exists
an optimal value for $d$. Figure~\ref{strain_drive} shows the
distribution of $\theta$ (since $\theta$ is periodic from zero to
one, it is only necessary to display the distribution of the decimal)
for different driving forces. As $d$ increases, the peak in the
distribution moves to a greater value and the whole distribution
flattens.

From these figures, it is clear that there is an optimal driving
force in which the melt-freeze phase exists. This behavior qualitatively
agrees with both the particle simulations and the one order parameter
phenomenological model in~\cite{Das1,Das2}.

\subsection{Coupling Dependence}

\begin{figure}
\begin{center}
\includegraphics[width=3in]{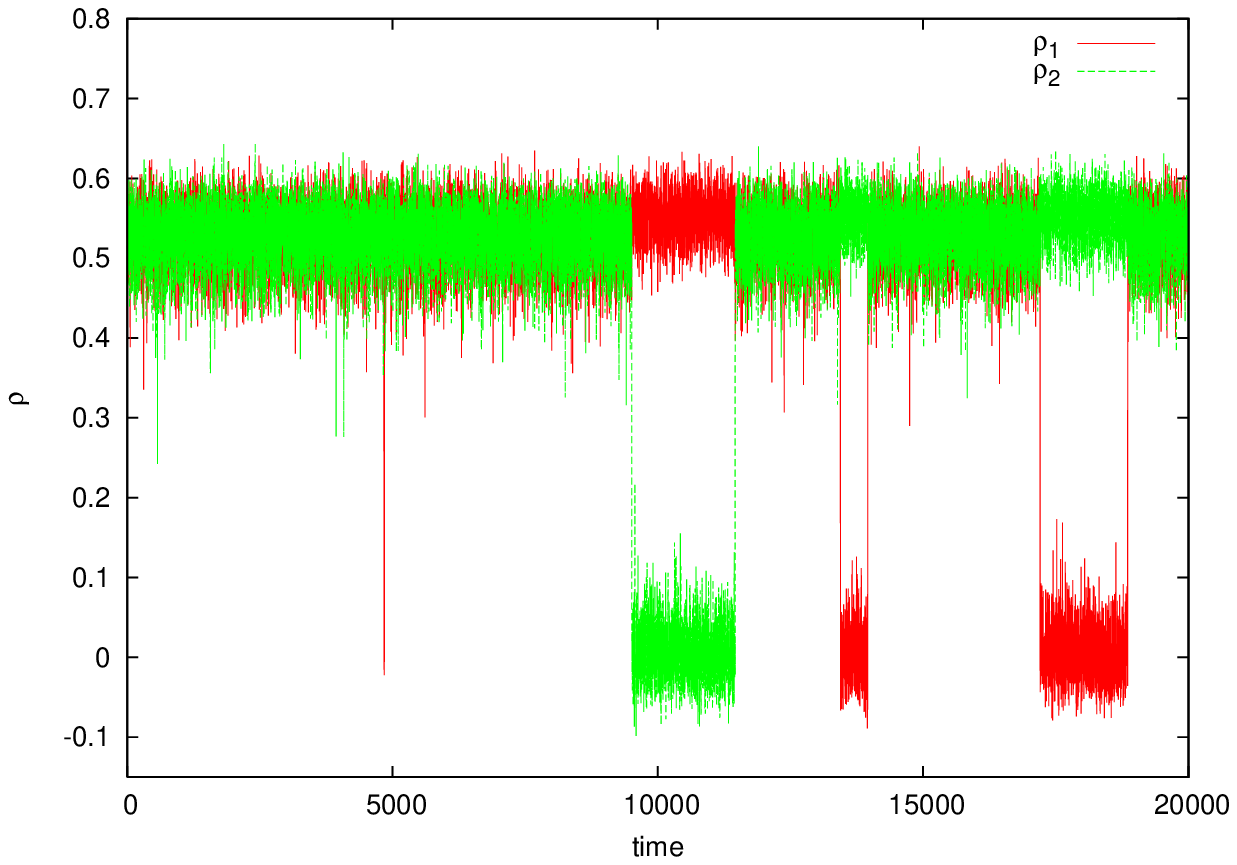} \\
\includegraphics[width=3in]{order_d0.1.eps} \\
\includegraphics[width=3in]{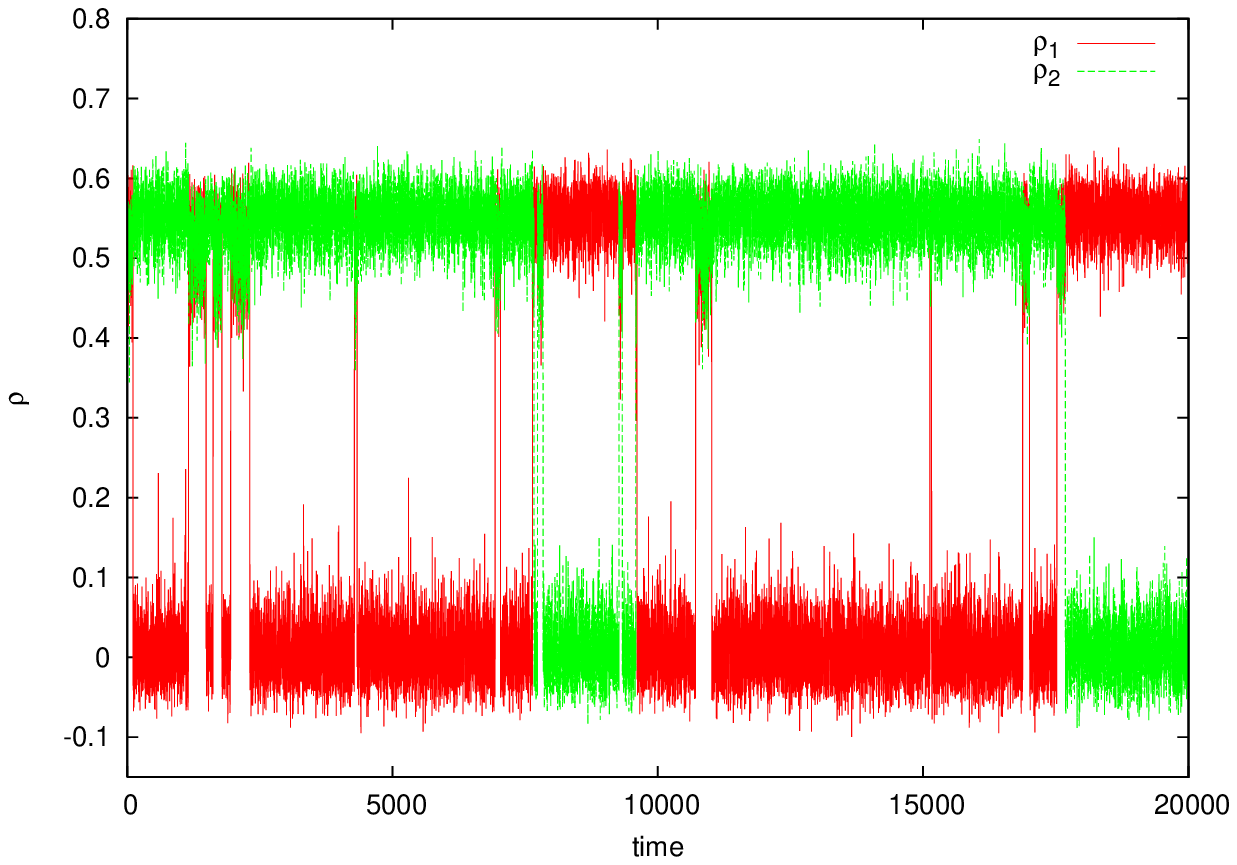} 
\caption{The order parameter vs. time for three different coupling
strengths $c = 0.2, 0.25,$ and $0.3$ respectively from top to
bottom. The driving force $d=0.1$ and noise strength $D=0.004$ are
kept fixed.}
\label{coup}
\end{center}
\end{figure}

\begin{figure}
\begin{center}
\includegraphics[width=3.25in]{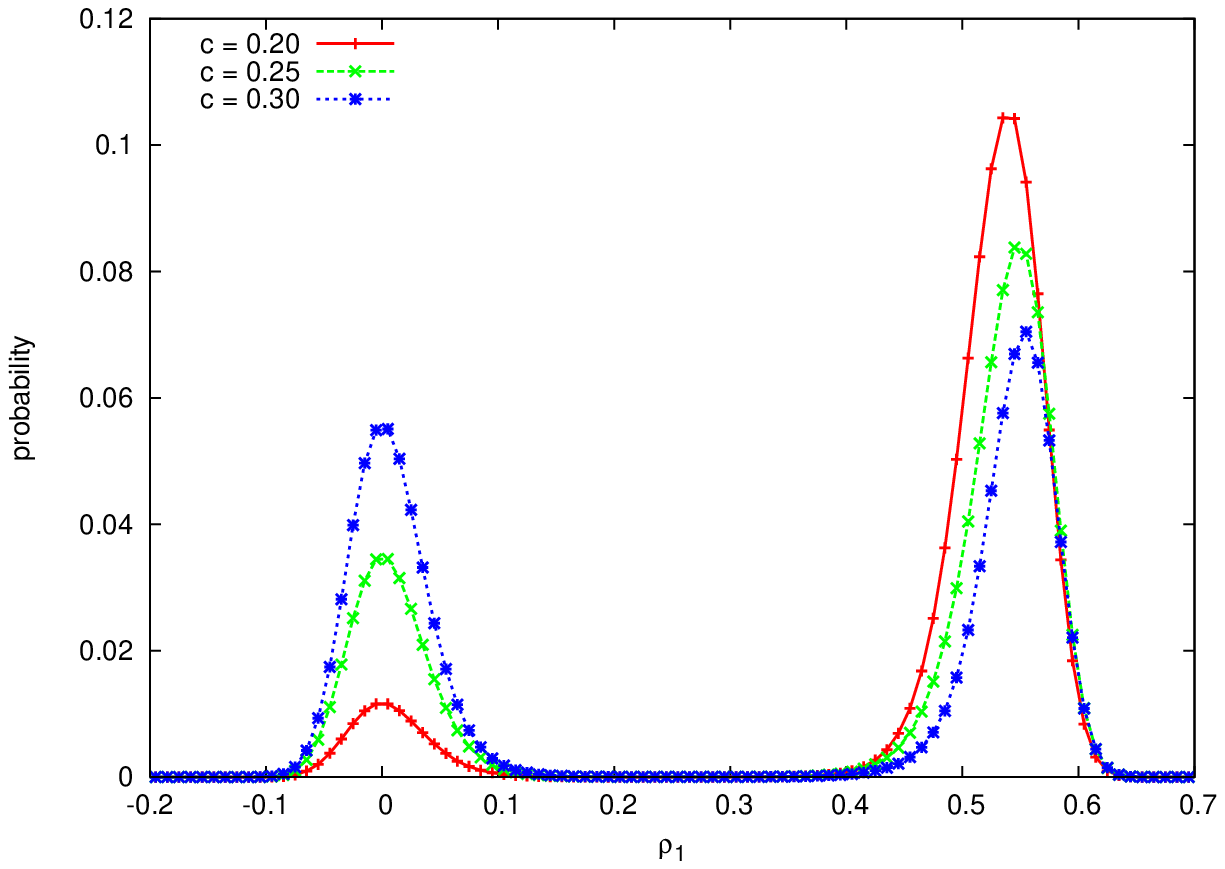} 
\caption{The distribution of $\rho_1$ taken over $10^{11}$
steps for $c = 0.2, 0.25,$ and $0.3$. The driving force $d=0.1$
and noise strength $D=0.004$ are kept fixed.}
\label{prob_coup}
\end{center}
\end{figure}

Keeping the driving force at a fixed value of $d = 0.1$ (which is
near the optimal value), we observe the behavior of the system as
the coupling constant varies. Figure~\ref{coup} shows the order
parameters versus time and figure~\ref{prob_coup} shows the
distribution of $\rho_1$ for different couplings.

Since the driving force is taken to be near the optimal value, all
three cases from figures~\ref{coup} and~\ref{prob_coup} are in the
melt-freeze phase. From these figures, it is clear that as the
coupling strength increases, the time spent in the liquid phase
increases as well, which is a feature seen in the models of Das\ea
Furthermore, systems with larger coupling strengths have order
parameters that tend to be in opposite phases.

To explore this further, we define the boundary between the two
phases to be the location of the local maximum without stress, $\rho
= (1/2\gamma) [\beta-(\beta^2-4\alpha\gamma)^{1/2}]$, and calculate
how often $\rho_1$ and $\rho_2$ are in the same well. In the weak
coupling case ($c = 0.2$), the two order parameters are in the same
well approximately 80 percent of the time. For the medium coupling
case ($c = 0.25$), $\rho_1$ and $\rho_2$ are in the same phase
approximately 35 percent of the time and in the large coupling case
($c = 0.3$), they only reside in the same well about 10 percent of
the time.  In the infinite coupling limit, the two order parameters
will ``repel'' each other rather strongly. Due to the symmetry in
the free energy between $\rho_1$ and $\rho_2$, in this limit, each
order parameter has the same probability of being in the ordered
phase as the disordered.

The observation described in the above paragraph provides a new
prediction for spatially dependent simulations and experiments.
Obviously, the behavior seen here cannot be seen in the one order
parameter mean field model.

By observing that the coupling term in the free energy merely changes
the coefficient of the quadratic term for a given order parameter,
one can intuitively understand this behavior.  In the large coupling
regime, the landscape in which $\rho_1$ exists is strongly determined
by $\rho_2$ and $\theta$. If both start in the crystal well, as
soon as $\theta$ is finite, this minimum quickly becomes unstable
and one of the $\rho$'s must roll down to the disordered minimum.
If $\rho_2$ falls down the liquid well, the amplitude of the effective
coupling drops near zero and $\rho_1$ is again happily in its stable
well. In addition, from $\rho_2$'s perspective, it is also in its
own stable well because $c\rho_1^2/2$ is finite and large. Since
the height of the cosine potential is proportional to both order
parameters squared, when one order parameter is in the liquid phase,
$\theta$ is essentially driven without resistance. Driving the
strain variable in this way allows the crystal well to become stable
for $\rho_2$. With aid of noise, $\rho_2$ can eventually move back
to the crystal phase. But once this occurs, both order parameters
are in a highly unstable position and one will quickly leave for
the liquid phase and the situation repeats itself. Which one crosses
to the liquid phase first is a stochastic process and cannot be
determined, but the repulsion effect still remains.

The simulations of layers of particles in~\cite{Das1,Das2} show
hints of this repulsive behavior though this repulsion is not as
explicitly seen as in our mean field model. Our phenomenological
model provides intuition into this behavior and predicts that it
should be more easily seen as the coupling increases. More work is
necessary to verify whether this is seen in a spatially dependent model.

With the form of our model free energy and the parameters in our
simulations, the repulsive behavior is so strong that the system
has yet to be seen in a state where both order parameters are in
the disordered phase. Though this has not been seen in our simulations,
the stochastic nature allows this state to be possible and, presumably,
parameters can be tuned so that both order parameters can be in
the liquid well in a quasi-stable state.

\subsection{Fluctuations}

A close inspection of figures~\ref{drive} and \ref{coup} reveals
interesting behavior regarding the fluctuations of the order
parameters; the fluctuation of one order parameter depends on the
state of the other order parameter. Obviously, the noise amplitude
and shape of the well determines the amount an order parameter
fluctuates. The magnitude of the other order parameter (and the
strain) influences the quadratic term of the free energy, thereby
changing the shape of the well.

Figures~\ref{drive} and \ref{coup} show that the fluctuations of
an order parameter in the crystalline well is much larger when the
other order parameter resides in the same well. To quantify this,
we compare the standard deviation of $\rho_1$ while both $\rho_1$
and $\rho_2$ are in the crystalline well ($\rho >
(1/2\gamma)[\beta-(\beta^2-4\alpha\gamma)^{1/2}]$) to the standard
deviation while $\rho_1$ is ordered and $\rho_2$ is disordered. For
$c=0.25, d=0.1,$ and $D=0.004$, the standard deviation is approximately
0.039 for the former and 0.028 for the latter.

This behavior makes physical sense because when both layers are
solid, stress is induced between the two layers and the particles
are driven further away from their equilibrium positions. Whereas,
if the particles in one layer are disordered, the layer will not
induce any stress on the neighboring layer.

\section{Discussion}

We note two minor disagreements between our mean field model and
the more detailed particle simulations of Das\ea\cite{Das1,Das2}.
The first is that Das\ea observed a smaller extent of order with
smaller interlayer coupling whereas we observe the opposite, \ie
for our model, the mean value for $\rho$ in the crystalline state
is larger for smaller coupling $c$. The second disagreement is that
for larger coupling we see larger fluctuations of the order parameter
while Das\ea observed a decrease in the amount of the fluctuations.
Considering the simplicity of the phenomenological model, disparities
such as these are expected and the many agreements are quite
surprising.

By varying the noise in the one order parameter phenomenological
model, Das\ea connects this melt-freeze phase to stochastic
resonance~\cite{Das1}. Qualitatively, the optimal driving force
exists when the driving time constant matches (within perhaps a
factor of two) the thermal barrier crossing time constant. Conversely,
with a fixed driving force, one expects a small window of noise
amplitudes where the thermal escape rate ``resonates'' with the
driving rate. The existence of an optimal noise amplitude~\cite{snr}
is a signature of stochastic resonance~\cite{mcnamara}.  Due to the
similarities of that model and ours, we expect that our model
displays stochastic resonance as well, though we have not performed
extensive simulations with noise dependence.

In this paper, we introduce a simple phenomenological model for a
system of two sheared monolayers to study a novel phase, the
melt-freeze phase, first seen in the simulations of
Das\ea~\cite{Das1,Das2}. Using numerical simulations, we observe
that the two order parameter model recovers many of the same
qualitative features as the simulations from Das\ea, especially the
dependence on driving force and coupling strength. Our simple model
displays new behavior involving the interplay of the order parameters
of two adjacent layers. One behavior is that with strong interlayer
couplings, two adjacent layers will be in opposite states of matter.
Another involves the fluctuation of the order parameters in the
melt-freeze phase; fluctuations are larger when both order parameters
coexist in the same well than when they are in opposite wells. More
simulations on spatially dependent systems are necessary to verify
these observations.

Das\ea suggested experiments using bulk colloidal crystals, ordered
copolymer monolayers, or colloidal monolayers to see this unusual
melt-freeze phase~\cite{Das1}. Systems of charged colloidal particles
are good candidates to observe this phenomenon because they possess
many properties that are tunable. The review article by
Palberg~\cite{palberg} thoroughly discusses the physical properties
of charged colloids, how to change the properties, and different
measurement techniques. For instance, the interparticle interaction
can be adjusted by changing the electrolyte concentration, thereby
altering the Coulomb screening. As seen in the simulations, the
melt-freeze phenomenon exists only in certain parameter windows,
therefore tunable parameters are necessary for experimental
observation. The melt-freeze phase and the behavior seen in our
mean field model could be seen in these experimental systems.

\section{Acknowledgments}
The author acknowledges Moumita Das and Sriram Ramaswamy for
informative discussions on their previous simulations. The author
also thanks Onuttom Narayan for useful discussions.


\begin{thebibliography}{}
\bibitem{order}
Actually, the order parameter cannot represent the structure factor
exactly due to its possible negativity, however, the qualitative
features are the same.
\bibitem{strain}
Our strain part of the free energy differs from the one used
in~\cite{Das1}. Both forms qualitatively incorporate the characteristics
of strain.
\bibitem{snr}
To be more precise, the signature of stochastic resonance is a peak
in the signal-to-noise ratio as a function of noise amplitude.
\end{thebibliography}
\end{document}